\documentclass[sigconf, nonacm]{acmart}
\AtBeginDocument{%
  }

\setcopyright{acmlicensed}
\copyrightyear{2018}
\acmYear{2018}
\acmDOI{XXXXXXX.XXXXXXX}
\acmConference[Conference acronym 'XX]{Make sure to enter the correct
  conference title from your rights confirmation email}{June 03--05,
  2018}{Woodstock, NY}
\acmISBN{978-1-4503-XXXX-X/2018/06}

\usepackage{float}
\floatstyle{ruled}
\newfloat{algorithm}{tbp}{loa}
\floatname{algorithm}{Algorithm}
\newcommand{\algI}{\hspace*{1.3em}}
\usepackage{tikz}

\usepackage[most]{tcolorbox}
\usepackage{marvosym}
\usepackage{booktabs}
\usepackage{multirow}
\usepackage{amsmath}

\definecolor{panelBack}{HTML}{F1F4F7}
\definecolor{panelRule}{RGB}{130,136,145}
\newtcolorbox{topmatterBox}{
    enhanced,
    colback=panelBack,
    colframe=panelBack,
    boxrule=0pt,
    frame hidden,
    arc=3mm,
    boxsep=0pt,
    left=6mm, right=6mm, top=5mm, bottom=4mm,
    before skip=4mm, after skip=0mm,
    notitle
}
\newcommand{\headerlogo}{%
  \includegraphics[trim={32.8 43.5 23.4 42.3}, clip, height=4mm]{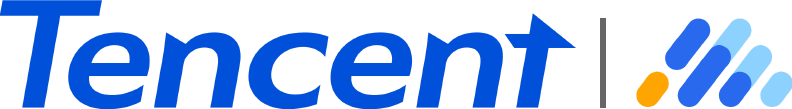}}

\newcommand{\panelnamefont}{\normalsize\bfseries}
\newcommand{\panelaffilfont}{\small}
\newcommand{\panelmailfont}{\ttfamily\footnotesize}

\makeatletter
\newbox\KF@titlebx
\let\@mktitle\@mktitle@i
\def\@titlefont{\Huge\bfseries}
\def\@mkauthors{\global\setbox\KF@titlebx=\box\mktitle@bx}
\def\@mkteasers{%
  \global\setbox\mktitle@bx=\vbox{%
    \hsize=\textwidth \linewidth=\textwidth \columnwidth=\textwidth
    \noindent\headerlogo\par
    \vskip 1.25mm                      
    {\color{panelRule}\hrule height 1.2pt}%
    \vskip 4.75mm                      
    \box\KF@titlebx
    \begin{topmatterBox}
      \setlength{\parindent}{0pt}%
      \authorpanel
      \medskip
      \noindent\ignorespaces\@abstract\par
      \ifx\@keywords\@empty\else
        \medskip\noindent{\bfseries\keywordsname:} \@keywords\par
      \fi
      \medskip
      {\color{panelRule}\hrule height 0.4pt}%
      \smallskip
      \noindent{\small\authornotepanel}\par
    \end{topmatterBox}
    \vskip 3.5mm                       
  }%
  \global\let\KF@concepts\@concepts
  \global\let\KF@keywords\@keywords
  \global\let\@mkabstract\@empty
  \global\@ACM@printccsfalse
  \global\let\@concepts\@empty
  \global\let\@keywords\@empty}
\let\KF@printendtopmatter\@printendtopmatter
\def\@printendtopmatter{%
  \global\let\@concepts\KF@concepts
  \global\let\@keywords\KF@keywords
  \hypersetup{pdfsubject={\@concepts}, pdfkeywords={\@keywords}}%
  \KF@printendtopmatter}
\makeatother

\newcommand{\authorpanel}{%
  {\panelnamefont
    Jiahao Hui$^{*}$,
    Lin Zhu$^{*}$\textsuperscript{\Letter},
    Ben Tan$^{*}$,
    Yishen Hu,
    Jingdong Shu,
    Zetai Jiang,
    Xining Ran,
    Yeshou Cai,
    Gong Chen,
    Haijie Gu\par}
  \smallskip
  {\panelaffilfont\raggedright
    \mbox{Tencent Inc., Shenzhen, China}\par}
  {\panelmailfont\raggedright
    \mbox{\{jiahaohui, techzhu, bentan, eastonhu, breezeshu, zetaijiang,}
    \mbox{xiningran, showcai, natchen, jerrickgu\}@tencent.com}\par}%
}
\newcommand{\authornotepanel}{%
  $^*$\,Equal contribution. \quad
  \textsuperscript{\Letter}\,Corresponding author.}

\renewcommand{\shortauthors}{Jiahao Hui, Lin Zhu}
\makeatletter
\gdef\authors{Jiahao Hui, Lin Zhu, Ben Tan, Yishen Hu, Jingdong Shu,
  Zetai Jiang, Xining Ran, Yeshou Cai, Gong Chen, Haijie Gu}
\makeatother

\begin{document}

\title[KuaFu: Compressing Long User Behavior into Understanding at Billion Scale]{KuaFu: Compressing Long User Behavior into Understanding at Billion Scale}
\begin{abstract}
Learning user preferences from long behavioral sequences is fundamental to large-scale advertising, recommender systems, and conversational agents. In industrial systems, preference modeling is typically task-wise, with different downstream tasks independently processing the same user history to capture different aspects of preference. This repeated processing creates substantial redundancy at billion-user scale, especially as preferences evolve and representations require frequent updates. We present \textbf{KuaFu}, a unified user preference modeling system that replaces repeated user-level sequence encoding with reusable item-level compressed representations. Each behavioral item is encoded once and shared across users and downstream tasks, while a user's history is constructed chronologically and extended with new item representations as behaviors arrive, enabling KuaFu to efficiently capture evolving preferences without re-encoding the entire history. To reduce both inference and storage overhead, KuaFu introduces a two-axis projector: token-axis compression shortens behavior sequences by up to 15$\times$ and reduces each item from 150~KB to 10~KB, while width-axis compression reduces the cached representation to 0.5~KB. We further develop a four-stage training pipeline consisting of reconstruction pre-training, preference-understanding post-training, joint compression-generation training, and hallucination-aware reinforcement learning, progressively improving representation fidelity and downstream utility. Across six MRQA and five RecBench public benchmarks, KuaFu consistently outperforms existing compression methods at comparable compression ratios, with gains of up to 17.7 EM and 1.90 AUC points respectively. KuaFu has been fully deployed and evaluated on \platform{} for ten months. Over this period, it improves per-GPU throughput by up to 350\%, saves 190 GPUs, and increases overall GMV by 1.37\%.
\end{abstract}



\settopmatter{printacmref=false}          
\renewcommand\footnotetextcopyrightpermission[1]{}  
\newcommand{\platform}{the Tencent advertising and recommendation platform}
\newcommand{\Platform}{The Tencent advertising and recommendation platform}







\maketitle

\section{Introduction}
\begin{figure*}[t]
  \centering
  \includegraphics[width=\textwidth]{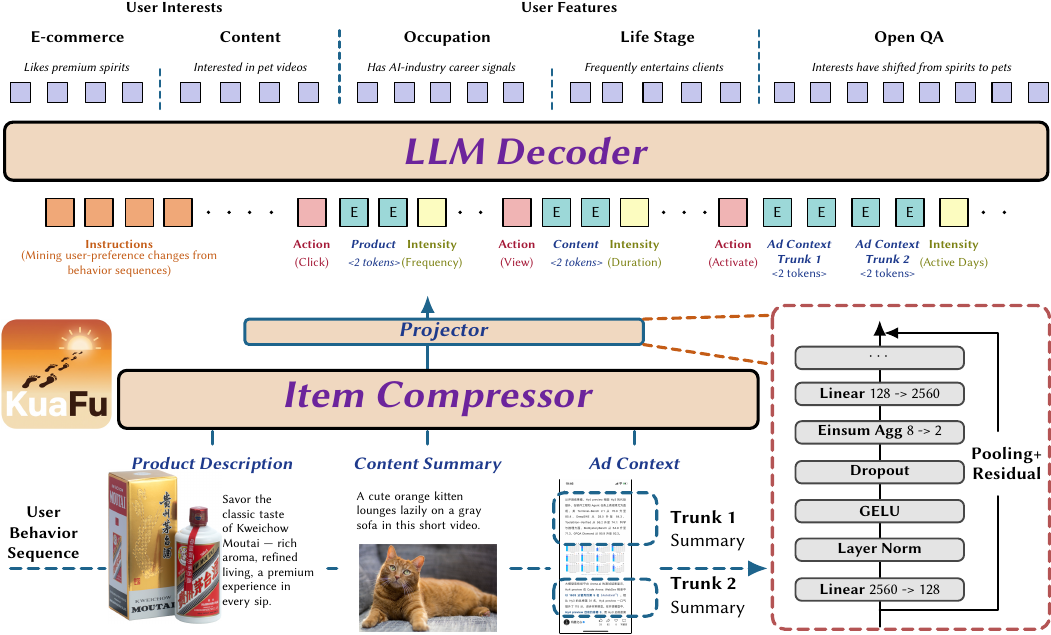}
  \caption{Overall architecture and workflow of KuaFu. Each behavioral item is independently encoded by the \emph{item compressor} and further reduced by the \emph{two-axis projector} into a small set of compact memory tokens. Items may correspond to heterogeneous behavioral content, such as product descriptions, content summaries, or advertisement context segments. For a given user, the resulting item representations are assembled in chronological order. The LLM decoder then consumes the assembled sequence to perform diverse user-understanding tasks, including interest modeling, user-feature inference, and open-ended question answering. Since each compressed representation depends only on the item itself, it can be cached offline and reused across users and downstream tasks, avoiding repeated encoding of long behavioral histories. }
  \label{fig:overview}
\end{figure*}

User behavioral sequences are a fundamental signal for large-scale advertising, recommender systems, and conversational agents, providing fine-grained evidence of evolving user preferences. In industrial systems, user preference modeling is typically \emph{task-wise}~\cite{appusage_survey}: different downstream tasks independently process the same behavioral history to infer aspects such as occupation, shopping interests, or content preferences. This repeated processing introduces substantial computational redundancy, especially at Web scale, where billions of users need to be served and frequently refreshed as their preferences evolve.

Recent advances in large language models (LLMs) offer a promising way to compress long behavioral histories into compact representations for downstream preference modeling~\cite{patchrec}. However, two challenges remain. First, \emph{aggressive compression can compromise preference fidelity}. In our production systems, we observe four recurring failure modes---\emph{fabrication}, \emph{omission}, \emph{temporal misattribution}, and \emph{broken reasoning}---which arise when important behavioral evidence is lost or distorted during compression. Prior work has similarly shown that prompt compression can increase hallucination~\cite{zhang2025hallucination}. Second, \emph{LLM-based long-sequence processing remains computationally expensive}: repeatedly encoding complete user histories for multiple downstream tasks and frequent updates incurs substantial GPU, storage, and serving overhead at industrial scale.

To address these challenges, we present \textbf{KuaFu}, a unified framework for efficient and fidelity-preserving user preference modeling at Web scale. As shown in Figure~\ref{fig:overview},KuaFu shifts the basic unit of representation from users and tasks to reusable behavioral items: each item is encoded into a compact representation that can be shared across users and downstream tasks, while a user's history is dynamically assembled from these representations in chronological order. Based on this item-centric abstraction, KuaFu jointly addresses representation efficiency and preference fidelity through a compact compression architecture, a progressive training strategy, and fidelity-oriented evaluation.

Our main contributions are summarized as follows:\\
\textbf{$\cdot$ An item-centric and cacheable architecture for Web-scale user modeling.}
 We propose a unified representation paradigm in which behavioral items are independently encoded by an item compressor and shared across users and downstream tasks. This design decouples expensive behavioral representation learning from task-specific preference modeling and supports incremental extensions as new behaviors arrive. A two-axis projector further compresses representations along both the token and feature dimensions, substantially reducing storage and serving costs. To the best of our knowledge, KuaFu is the first system to encode behavioral text at item granularity and deploy such representations in a production environment at billion-user scale.
\\
\textbf{$\cdot$ A fidelity-oriented four-stage training pipeline.}
We construct a progressively challenging training corpus that spans reconstruction, behavior understanding, long-sequence reasoning, and downstream user-understanding tasks. Building on this curriculum, KuaFu is trained in four stages: reconstruction pre-training, preference-understanding post-training, joint compression-generation training, and hallucination-aware reinforcement learning. This staged design progressively improves information preservation, semantic understanding, generation quality, and robustness to compression-induced hallucinations. A sequence-length curriculum further stabilizes training on long behavioral sequences, while the reinforcement-learning objective explicitly penalizes different hallucination modes and rewards completeness.
\\
\textbf{$\cdot$ A layered evaluation protocol with large-scale production validation.}
We evaluate compressed representations directly across tasks of increasing difficulty, from basic classification and multi-label preference attribution to question answering and multi-task transfer, providing a shorter feedback loop between compression quality and downstream performance. Beyond offline evaluation, KuaFu has been fully deployed on \platform{} at billion-user scale. In production, it reduces each cached item representation from 10~KB to 0.5~KB, improves per-GPU throughput by up to 350\%, saves 190 GPUs, and delivers a 1.37\% improvement in overall GMV over ten months of deployment.

\section{Related Work}
\label{sec:related}
\textbf{Generic context compression.}
Existing context-compression methods can be broadly divided into hard and soft compression. Hard-compression methods, such as Selective Context~\cite{selectivecontext} and LLMLingua~\cite{llmlingua}, remove less informative tokens while preserving natural-language inputs. Soft-compression methods instead map context into compact continuous representations, including GIST~\cite{gist}, AutoCompressor~\cite{autocompressor}, ICAE~\cite{icae}, 500xCompressor~\cite{x500compressor}, REFRAG~\cite{refrag}, SAC~\cite{sac}, and EPL~\cite{epl}. These methods demonstrate that substantial compression can preserve downstream QA performance, but they primarily target documents or retrieved passages. In contrast, KuaFu compresses timestamped behavioral items into reusable representations and explicitly optimizes fidelity for user-preference understanding, including hallucination patterns such as fabrication, omission, and temporal misattribution~\cite{zhang2025hallucination}.

\textbf{Item and behavior representations in recommendation.}
Generative recommender systems increasingly represent items or behavioral histories with compact discrete or continuous tokens. Representative approaches include semantic-ID methods such as TIGER~\cite{tiger}, multi-facet item representations such as LAMIA~\cite{lamia}, heterogeneous feature tokenization in Token Factory~\cite{tokenfactory}, session-level compression in PatchRec~\cite{patchrec}, and query-based bottlenecks such as Q-Former~\cite{qformer} and U$^2$QT~\cite{u2qt}. These methods are effective for recommendation objectives such as HR, NDCG, and CTR, but they typically evaluate representations only through end-to-end recommendation quality. KuaFu instead focuses on whether compressed behavioral representations preserve user-preference evidence and temporal structure across heterogeneous downstream tasks, including profiling and conversational user understanding. Recent benchmarks such as GISTBench~\cite{gistbench} and UserSumBench~\cite{usersumbench} move toward verifiable preference evaluation, but do not jointly study compression ratio, representation fidelity, and downstream profiling quality.

\textbf{Long-sequence modeling and LLM efficiency in industrial systems.}
Industrial systems address long user histories and LLM serving cost from several directions. Long-sequence recommendation methods directly model extended behavior histories through large-scale sequence modeling or sparse attention~\cite{bytedance_makeitlong,meituan_sparseattn}, while LLM-efficiency work reduces serving cost through distillation, latent reasoning, KV-cache pruning, or faster decoding~\cite{meituan_distill,alibaba_cotlatent,baidu_probefetch,alibaba_nezha}. Generative recommendation systems also exploit pretrained language models and semantic item IDs in production~\cite{youtube_plum,taotian_lsig}. KuaFu is complementary to these approaches: rather than modifying the ranking model or replacing the LLM, it compresses behavioral inputs into item-level representations that can be cached offline and reused across users, requests, and downstream tasks.

\section{Problem Definition}
\label{sec:problem}
Let the chronological behavioral sequence of a user $u$ be $X_u = \{x_1^u, x_2^u, \ldots, x_{N_u}^u\}$, where $N_u$ denotes the length of the sequence and $x_i^u$ is the $i$th user action. Each action is represented as $x_i^u = (a_i^u, p_i^u)$, where $a_i^u \in \mathcal{A}$ denotes the action type, such as click, view, or read, and $p_i^u \in \mathcal{P}$ denotes the item involved in the action. The item pool $\mathcal{P}$ may contain heterogeneous content, including videos, advertisements, images, products, or document segments.
In conventional user modeling systems, different downstream tasks independently process $X_u$ to derive task-specific user representations. As a result, the same long behavioral sequence may be repeatedly encoded by multiple tasks, leading to substantial redundant computation. KuaFu instead adopts an item-centric representation paradigm. Given behavioral sequences collected from users, KuaFu learns a shared representation $\mathbf{z}_k = f(p_k; \mathcal{X})$, for each item $p_k \in \mathcal{P}$, where $\mathcal{X}={X_u}_{u=1}^{M}$ denotes the collection of user behavior sequences and $f(\cdot)$ is the item representation model. The representation $\mathbf{z}_k$ captures behavioral information associated with item $p_k$ and is shared across all users and downstream tasks. When a new interaction involving $p_k$ arrives, KuaFu incrementally extends $\mathbf{z}_k$ using the newly observed behavioral context. Therefore, an arbitrary user sequence can be represented by retrieving the corresponding item embeddings and preserving their chronological order:
$ Z_u = \{\mathbf{z}_{p_1^u}, \mathbf{z}_{p_2^u}, \ldots, \mathbf{z}_{p_{N_u}^u}\}$.
A downstream task $t$ can then construct its task-specific user representation from $Z_u$, $\mathbf{h}_u^{(t)} = g_t(Z_u)$, where $g_t(\cdot)$ is a lightweight task-specific preference function, such as user attribute prediction, shopping preference summarization, personalized question answering.

\section{The Proposed Framework}
\label{sec:method}


\begin{figure*}[t]
  \centering
  \includegraphics[width=0.8\textwidth]{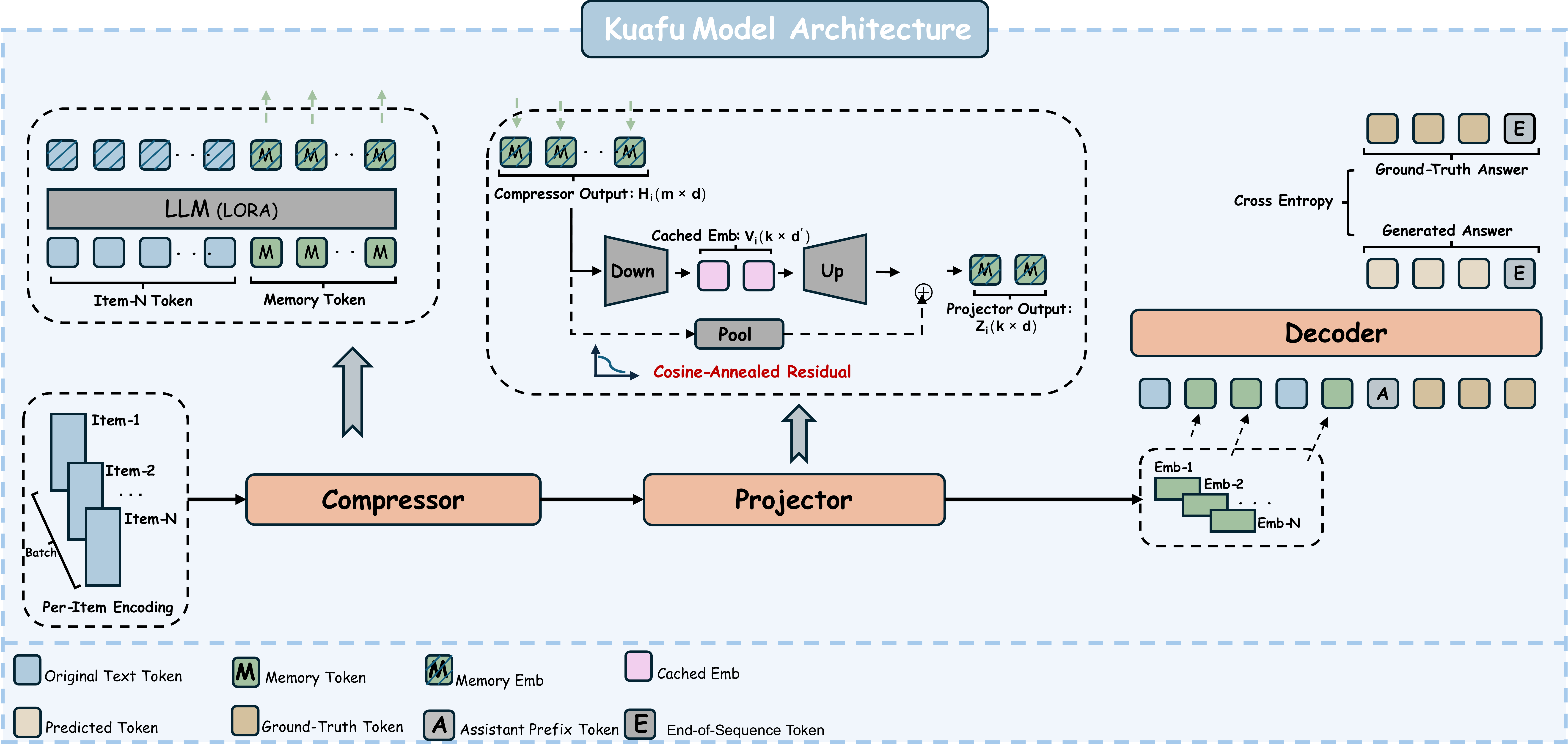}
  \caption{The architecture of the proposed KuaFu framework. }
  
  \label{fig:arch}
\end{figure*}

In this section, we describe the architecture and training pipeline of KuaFu. As illustrated in Figure \ref{fig:arch}, KuaFu consists of three main components: a \emph{compressor}, a \emph{projector}, and a \emph{decoder}. The compressor extracts compact semantic representations for individual items, the projector further reduces their storage, and the decoder consumes the resulting representations to perform downstream user modeling and generation.

To obtain compact representations that retain sufficient information for downstream generation, we adopt a four-stage training pipeline: (1) reconstruction pre-training, (2) preference understanding post-training, (3) joint compression-generation training, and (4) hallucination-aware reinforcement learning. These stages progressively improve the semantic fidelity and downstream utility of the compressed representations.

\subsection{Main Components}
\paragraph{Compressor}
The compressor $E$ is a causal language model with hidden dimension $d$, adapted using LoRA. For an item $p_i$, let $\mathbf{S}_i = (s_i^1, \ldots, s_i^{L_i})$ denote the token sequence obtained from its textual description, where $L_i$ is the number of text tokens. We append $m$ learnable memory tokens to $\mathbf{S}_i$. These memory tokens use a dedicated embedding table rather than the model's ordinary vocabulary, while the text tokens retain their original input embeddings. A single forward pass of the compressor produces the final-layer hidden states at the memory-token positions,
\begin{equation}
  \mathbf{h}_i \;=\; \mathrm{Enc}_E\!\left( \mathbf{S}_i \oplus [\textsc{mem}]_{1:m}\right)
  \;\in\; \mathbb{R}^{m\times d},
  \label{eq:enc}
\end{equation}

Unlike conventional user-centric approaches that repeatedly encode long user histories, KuaFu processes each item independently. As a result, item representations can be cached and reused across different users and downstream tasks. 

\paragraph{Projector}
To further reduce the representation size, as illustrated in the middle of Figure~\ref{fig:arch}, we introduce a \emph{two-axis projector} that decreases both the number of vectors and their feature(width) dimension. For example, the projector maps the $m(=8)$ memory vectors to $2$ and the width from $d(=2{,}560)$ to $128$. 

Specifically, in the projector, a $Down$ network, $Down(\mathbf{h}_i)\in \mathbb{R}^{m\times d'}$, maps $m$ vectors to width $d'$, then a learnable combination matrix $\mathbf{C}\in\mathbb{R}^{k\times m}$ transform $m$ vectors to $k$. In order to fit the input of the LLM decoder in the next component, we also have a $Up$ network that maps the $d'$-dimension vectors back to $d$ with a residual pooling function. Overall, the function of the projector is summarized as follows:
\begin{equation}
\tilde{z}_i \;=\; C\cdot \textsc{Down}(h_i), \qquad
z_i \;=\; \textsc{Up}\!\left(\tilde{z}_i\right) \;+\; \alpha(t)\,\mathrm{Pool}(h_i).
\label{eq:proj}
\end{equation}

Here $\alpha(t)$ is a cosine-annealed residual coefficient that is active only
over the first third of training steps and vanishes thereafter,
\begin{equation}
\alpha(t) \;=\;
\begin{cases}
  \tfrac{1}{2}\!\left(1 + \cos\dfrac{\pi t}{T_0}\right), & t \le T_0,\\[4pt]
  0, & t > T_0,
\end{cases}
\qquad T_0 = \tfrac{1}{3}T,
\label{eq:anneal}
\end{equation}
where $t$ is the training step and $T$ the total number of steps, so the pooled
residual decays smoothly from $1$ to $0$ and is a transient training-time
stabilizer rather than part of the converged model. The detailed architecture of the projector is shown at the bottom right of Figure~\ref{fig:overview}.

After the first third of training, $\alpha(t)$ decays to zero, so the converged representation becomes $z_i = \textsc{Up}(\tilde{z}_i)$ and no longer depends on the full hidden representation $h_i$. At serving time, KuaFu therefore caches only the low-dimensional core $\tilde{z}_i$ and
reconstructs the decoder input through the lightweight \textsc{Up} projection;
the pooled residual is used only during early training to stabilize optimization and is removed from inference. In the
production system, this two-axis compression substantially reduces the storage
footprint of item representations. Relative to the original $m \times d$ memory
representation, the cache holds a $k \times d'$ vectors of $\tilde{z}_i$, the compression ratio is
$\frac{m}{k}\cdot\frac{d}{d'}$. For example, in production, we have $m \times d=8\times 2560$\ and $k \times d'=2\times128$ floats, which are about 40 and 0.5\,KB respectively, the ratio is 80. From the original user behavior sequences, writing $L_i$ (=30, in average) for the raw token count of $x_i$, the effective sequence length seen by the decoder shrinks from
$L_i$ text tokens to $k$ compressed vectors, so the token-axis ratio is $L_i/k$
and the width-axis ratio is $d/d'$, the final ratio is up to $\frac{30\times 2560}{2\times 128}=300$.

\paragraph{Decoder}
As seen at the top right corner of Figure~\ref{fig:arch}, the decoder $D$ consumes item representations alongside ordinary text without extending its vocabulary. Each item reference in the input is replaced by its projected vectors $\mathbf{z}_i$, while text/instruction tokens retain the decoder's standard word embeddings. The decoder generates a response autoregressively from the resulting sequence. Both models use one-dimensional RoPE: the compressor encodes positions within each item input, including its memory tokens, and the decoder encodes positions across the assembled input. Concatenating item representations in chronological order therefore preserves the order of user interactions.
Algorithm~\ref{alg:kuafu} in the appendix lays out the full batched forward
pass used in training.

\subsection{Four-Stage Training}
\begin{table}[t]
  \centering
  \caption{Trainable set per stage. C/P/D denote compressor, projector
  and decoder respectively.
  }
  \label{tab:stages}
  \footnotesize
  \setlength{\tabcolsep}{4pt}
  \begin{tabular}{llccc}
    \toprule
    Stage & Objective & C & P & D \\
    \midrule
    Pre-training & reconstruct & update & --- & freeze \\
    Post-training & compressed QA & update & update & freeze \\
    Joint-training & fit both ends & update & update & update \\
    Hallu.-aware RL & de-hallucinate & freeze & freeze & update \\
    \bottomrule
  \end{tabular}
\end{table}

To learn high-quality compressed representations, KuaFu adopts a four-stage training pipeline: three supervised stages followed by hallucination-aware reinforcement learning. As described in Table~\ref{tab:stages}, the supervised stages share the same data flow, i.e., items are encoded independently, ordered chronologically, and fed to the decoder, but differ in their objectives, trainable modules, and use of the projector. Each stage is initialized from the previous one and follows a sequence-length curriculum from short to long histories. Together, the four stages progressively improve information preservation and grounded generation. In Section~\ref{sec:needle}, we also evaluates representation fidelity using a needle-in-a-haystack probe and quantifies the gains introduced by each training stage. Here, we introduce the details of each stage as follows.

\paragraph{Reconstruction Pre-training.}
We first train the compressor to retain enough information to reconstruct its input. The compressor maps each item input $\mathbf{S}_i$ to memory states $\mathbf{h}_i$, which are passed directly to the decoder with the projector disabled. The decoder reconstructs the original item inputs from the chronologically concatenated memory states, requiring it to recover each item in the presence of other compressed items. Only the compressor's LoRA parameters and memory-token embeddings are updated; the decoder remains frozen but provides the reconstruction loss. Because training directly on long histories fails to converge, we increase the number of items per sample across five buckets: $1$, $1$--$3$, $1$--$200$, $1$--$500$, and $1$--$1200$. The objective function is as follows:
\begin{equation}
\mathcal{L}_{\mathrm{AE}}=\max_{\varphi,\,e_m}\ \sum_{t=1}^{L_i}\log D_{\theta}\bigl(s_t\mid s_{<t},\,\mathbf{h}_1,\ldots,\mathbf{h}_{L_i}\bigr)
\end{equation}

\paragraph{Preference Understanding Post-training.}
Next, we attach the projector and train the system to answer questions about the compressed history. The decoder remains frozen, while the compressor's LoRA parameters, memory-token embeddings, and projector are updated. This stage shifts the learning objective from reconstructing each item to preserving the evidence needed for question answering: the decoder must retrieve relevant information from multiple item representations and combine it to answer a question about the user's history. 

\paragraph{Joint Compression-Generation Training.}
We then jointly train the compressor, projector, and decoder on question answering. Unlike the preceding stage, the decoder is now free to adapt to the compressed representations, while the compressor and projector can shape those representations around the information the decoder needs. This joint optimization improves the interface between compression and generation. 

\paragraph{Hallucination-aware reinforcement learning.}
Finally, we optimize the decoder to produce faithful and sufficiently complete answers from cached item representations. The compressor and projector are frozen, and item embeddings are loaded from the offline cache. For each prompt, we sample a group of $G$ answers and use a generative reward model to assess each answer against the original item text. The reward penalizes four error types---fabrication, missed detection, date misattribution, and faulty reasoning---and adds a completeness bonus so that the decoder cannot avoid errors merely by withholding information:

\begin{equation}
  r(y) \;=\; b\!\left(\mathrm{comp}(y)\right)
  \;-\; \sum_{c=1}^{4} w_c \cdot \mathbb{1}\!\left[\text{$y$ hits error $c$}\right].
  \label{eq:reward}
\end{equation}
The $G$ answers in a group share one prompt and the advantage is normalized by
the group mean. The policy is updated with DAPO~\cite{dapo}: token-level
importance ratio, asymmetric clipping, no KL term, and truncated answers
excluded from the loss, which mitigates the length collapse of
GRPO~\cite{grpo}.


\section{Experiments}
\label{sec:exp}
\subsection{Setup}
We introduce the experimental setup of models, training data, baseline, evaluation and detailed implementation as follows:

\textbf{Models.}
KuaFu uses an LLM encoder--decoder architecture, with Qwen3-4B~\cite{qwen3} as the default backbone for both components. For fair comparison with external compression baselines, we additionally use Llama-3.2-1B~\cite{llama32}. We also validate KuaFu on Qwen3-VL-4B-Instruct with a frozen ViT encoder.

\textbf{Training data.}
Training data combine large-scale advertising behavior sequences with general-purpose text and NLP corpora. After filtering and deduplication, samples are organized into compression--reconstruction and compression--understanding tasks with increasing difficulty. The final corpus covers 42 task types, from basic reconstruction to long-sequence and downstream user-understanding tasks. Detailed statistics are provided in Appendix~A.

\textbf{Baselines and evaluation.}
We compare against an uncompressed multi-task LLM, representative text-compression methods, and LLM-based recommendation models. Evaluation includes in-house user-understanding tasks, MRQA~\cite{mrqa}, RecBench~\cite{recbench}, and online production experiments. We report task accuracy metrics together with serving efficiency and GMV.

\textbf{Implementation.}
Models are trained for 3--5 epochs with learning rates of \(10^{-5}\) to \(10^{-4}\); the reinforcement-learning stage uses DAPO. Production inference is implemented with vLLM~\cite{vllm}.

\subsection{Production Profiling Tasks}

KuaFu is deployed in the profiling pipeline of \platform{}, where it processes weekly behavior histories for over one billion users and produces structured features for recommendation. At this scale, throughput is a primary bottleneck, motivating item-level compression. 

The production tasks fall into two families. \emph{Interest summarization} covers content and e-commerce interests, extracting verifiable preference signals from hundreds to thousands of cross-modal behaviors with a compression ratio of $10\times$. \emph{User profiling} covers relatively stable attributes such as occupation, industry, and life stage, using shorter behavior histories with a compression ratio of $15\times$. Both families share the same item-level compressed representations and differ only in their downstream prediction heads and evaluation metrics.

Table~\ref{tab:online} compares the previous uncompressed multi-task model with KuaFu under the same production setting. Interest summarization is evaluated by F1, while profiling tasks use classification accuracy. KuaFu matches or improves upon the uncompressed model on all five metrics, with gains ranging from $+0.4$ to $+2.0$ points. In addition to enabling weekly refreshes at billion-user scale, compression may also suppress redundant or weakly related behaviors, making stable preference signals easier to extract.

\begin{table}[t]
  \centering
  \caption{Downstream user-understanding tasks in production. $\Delta$ is the
  improvement in percentage compared with the multi-task online
  model.}
  \label{tab:online}
  \footnotesize
  \setlength{\tabcolsep}{3pt}
  \begin{tabular}{llccccc}
    \toprule
    Task & Family & items & ratio & online & KuaFu & $\Delta$ \\
    \midrule
    Content interest & summary & 400/1200 & $10\times$ & 85.9 & \textbf{86.5} & $+0.6$ \\
    E-com. interest & summary & 150/200 & $10\times$ & 95.7 & \textbf{96.1} & $+0.4$ \\
    Industry & profile & 30/100 & $15\times$ & 81.0 & \textbf{82.1} & $+1.1$ \\
    Occupation & profile & 30/100 & $15\times$ & 86.7 & \textbf{87.6} & $+0.9$ \\
    Life stage & profile & 16/100 & $15\times$ & 76.0 & \textbf{78.0} & $+2.0$ \\
    \bottomrule
  \end{tabular}
\end{table}

\subsection{In-House Benchmark against External Compressors}
\begin{table}[t]
  \centering
  \caption{In-house benchmark (percentage scale). All soft baselines are
  retrained under the aligned protocol described in the text. $\dagger$: training-free hard
  compression at $5\times$, listed as an external reference only.}
  \label{tab:inhouse}
  \footnotesize
  \setlength{\tabcolsep}{3pt}
  \begin{tabular}{lccccc}
    \toprule
    Method & E-com. & Occup. & Content & Life & Overall \\
    \midrule
    LLMLingua-2$^{\dagger}$ & 49.7 & 31.1 & 22.9 & 22.7 & 31.6 \\
    ICAE                          & 50.5 & 43.1 & 30.9 & 21.7 & 36.6 \\
    SAC                           & 48.3 & 50.7 & 34.3 & 23.7 & 39.2 \\
    500xCompressor                & 55.7 & 46.1 & 32.0 & 23.9 & 39.4 \\
    EPL                           & 53.9 & 56.3 & 34.7 & 24.5 & 42.4 \\
    \textbf{KuaFu}                & \textbf{71.7} & \textbf{64.5} & \textbf{64.7} & \textbf{51.1} & \textbf{63.0} \\
    \midrule
    $\Delta$ vs.\ best baseline   & $+16.0$ & $+8.2$ & $+30.0$ & $+26.6$ & $+20.6$ \\
    \bottomrule
  \end{tabular}
\end{table}

We evaluate KuaFu against representative compression methods on four in-house downstream user-understanding tasks. To ensure a fair comparison, all soft-compression baselines are retrained using the same behavior sequences, train/validation split, downstream supervision, input format, Llama-3.2-1B backbone, and 15$\times$ compression ratio. Each method follows its original training objective and is trained to convergence. LLMLingua-2, a training-free hard-compression method, is evaluated separately using its official checkpoint at 5$\times$ compression. All results are scored by an LLM judge.

From the result in Table~\ref{tab:inhouse}, we can see that KuaFu achieves the best overall score of 63.0 and outperforms the baseline on the four tasks by 16.0, 8.2, 30.0, and 26.6 points, respectively. No baseline exceeds 60 on any task, while KuaFu consistently performs best across all four. Since the major experimental factors are controlled, these gains highlight the advantage of KuaFu's item-level compression design, which represents each behavioral item with a compact group of embeddings rather than compressing an entire sequence into a small number of tokens.

\subsection{Generic Compression: MRQA In-Domain and Out-of-Domain}
\begin{figure}[tb]
  \centering
  \includegraphics{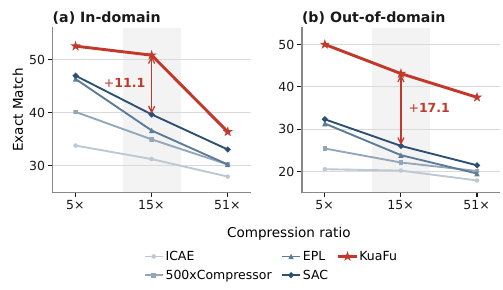}
  \caption{MRQA exact match vs.\ compression ratio (Llama-3.2-1B). Shaded
  band: $15\times$, the ratio used everywhere else. ROUGE-1 F1 and all baseline
  values: Appendix~\ref{app:mrqa}.}
  \label{fig:mrqa}
\end{figure}



To evaluate generic compression ability, we retrain KuaFu on SAC's training data following the same protocol, while baseline results are taken from the SAC paper. We evaluate both in-domain performance on six MRQA subsets and zero-shot out-of-domain transfer using the same checkpoint without adaptation.

As shown in Figure~\ref{fig:mrqa}, KuaFu consistently outperforms existing methods across compression ratios. At 15$\times$ compression, KuaFu achieves 50.8 EM in-domain, already surpassing SAC at 5$\times$ level (46.95), while at 51$\times$ it remains stronger than all baselines evaluated at 5$\times$ in the out-of-domain setting. These results demonstrate strong compression efficiency and cross-domain generalization.

\subsection{Recommendation Benchmark}
We further evaluate whether the compressed memory preserves recommendation signals using RecBench's zero-shot pairwise CTR setting across five datasets: MIND, MicroLens, Goodreads, CDs, and H\&M. Without any recommendation-specific training, KuaFu zero-shot achieves an average score of 0.617, outperforming most zero-shot baselines.

More importantly, KuaFu zero-shot improves over the same Qwen3-4B backbone from 0.570 to 0.617, a gain of 4.70 points, indicating that the improvement comes from the compressed memory representation rather than model scale. It also surpasses Qwen3-8B (0.598) by 1.90 points. The only exception is the CDs subset, where KuaFu zero-shot underperforms the Qwen3-4B backbone (0.489 vs.\ 0.510). Results for all zero-shot baselines are reported in Appendix~\ref{app:recbench}.


\subsection{Hallucination-Aware Reinforcement Learning}
\begin{figure}[tb]
  \centering
  \includegraphics{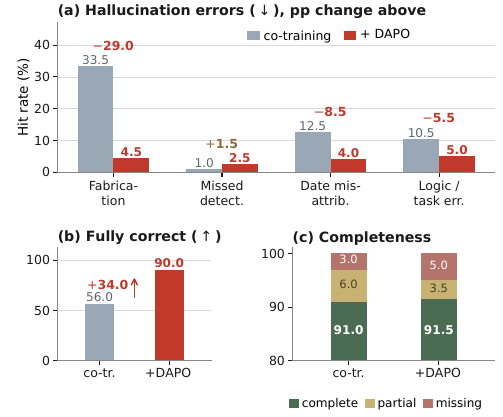}
  \caption{Hallucination-aware RL, scored by GPT-5.5, a stronger LLM than the
  training reward model; agreement with human raters and a second judge is in
  Appendix~\ref{app:judge}. (a) hit rate per error type (lower better), pp change above
  each pair; (b) fully correct answers; (c) completeness, shares summing to
  $100\%$. All values in percent.}
  \label{fig:rl}
\end{figure}

To improve open-ended reasoning over compressed long behavior sequences, we apply hallucination-aware DAPO starting from the joint training checkpoint. The compressor is frozen and only the decoder is updated. The reward model evaluates four hallucination types, i.e., fabrication, omission, temporal misattribution, and broken reasoning, together with answer completeness against the original uncompressed text.

We evaluate on 1,000 long-sequence open-ended QA samples using a stronger, independent LLM judge. As shown in Figure~\ref{fig:rl}, reinforcement learning substantially reduces hallucination: fabrication drops from 33.5\% to 4.5\%, while the fully correct rate increases from 56.0\% to 90.0\%. Other error types decrease similarly, with only a small increase in missed or incomplete answers. Overall completeness remains nearly unchanged at 91.5\%, indicating that the model becomes more conservative without simply refusing to answer. Human evaluation and an additional LLM judge show consistent trends, with judge--human agreement of ($\kappa=0.72$)--(0.91). More details are shown in Table~\ref{tab:judge-agree} in Appendix~\ref{app:judge}.

\subsection{Retrieval Fidelity Across Training Stages}
\label{sec:needle}
\begin{figure}[t]
  \centering
  \includegraphics{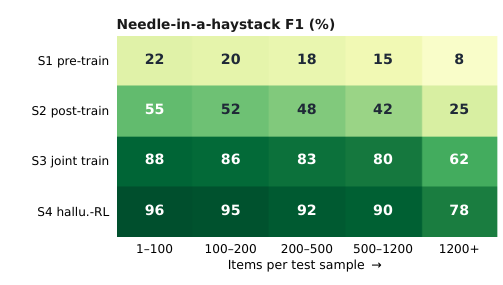}
  \caption{Needle-in-a-haystack retrieval fidelity (F1) across the four training
  stages (rows) by items per test sample (columns); greener (darker) shades are
  higher.}
  \label{fig:needle}
\end{figure}

Figure~\ref{fig:needle} evaluates how well compressed representations preserve item-level information using a needle-in-a-haystack retrieval probe. Relevant items are hidden within long behavior sequences, and the model must recover them from the compressed input; retrieval F1 is reported across different sequence lengths and training stages.

Two trends are clear. First, retrieval fidelity improves consistently across the four training stages, rising from 8--22 after pre-training to 78--96 after hallucination-aware reinforcement learning. Second, performance decreases as sequence length grows, but degrades gracefully rather than collapsing: the final model still achieves 78 F1 on sequences with more than 1,200 items. These results show that the four-stage training pipeline progressively improves information preservation, including under very long-context settings.


\subsection{Ablations}

\begin{table}[t]
  \centering
  \caption{Ablation summary. Best value of each group is in bold; per-group
  details are in Appendix~\ref{app:ablation}.}
  \label{tab:ablation}
  \footnotesize
  \setlength{\tabcolsep}{4pt}
  \renewcommand{\arraystretch}{1.08}
  \begin{tabular}{p{0.28\columnwidth} p{0.64\columnwidth}}
    \toprule
    Group & Details \\
    \midrule

    Curriculum
      & \textit{Setting:} mixed $1$--$1200$ (980K) vs.\ five increasing buckets \\
      & \textit{Metric:} train loss 2.27 vs.\
        \textbf{0.2, 0.28, 0.16, 0.02, 0.034} \\
    \midrule

    Inheritance
      & \textit{Setting:} without vs.\ with pre-training \\
      & \textit{Metric:} init/train/val loss
        3.2/0.7/4.1 vs.\ 1.2/0.6/\textbf{0.8} \\
    \midrule

    Token axis
      & \textit{Setting:} pooling / adapter / adapter $+$ pooled residual \\
      & \textit{Metric:} train loss
        1.9967 / 1.8026 / \textbf{1.4148} \\
    \midrule

    Width axis
      & \textit{Setting:} naive / $+$residual / $+$LR annealing /
        $+$continued training \\
      & \textit{Metric:} train reward
        0.1 / 0.25 / 0.6 / \textbf{0.8} \\
    \midrule

    Decoder
      & \textit{Setting:} Base / SFT / DAPO \\
      & \textit{Metric:} life stage 0.74 / 0.81 / 0.81;
        e-commerce 0.806 / 0.823 / \textbf{0.826} \\

    \bottomrule
  \end{tabular}
\end{table}

We ablate four design choices across KuaFu: curriculum construction and weight inheritance for the compressor, token- and width-axis design for the projector, and backbone training strategy for the decoder. Table~\ref{tab:ablation} shows that each component contributes to optimization stability or downstream performance, while progressively combining the proposed designs generally yields stronger results than simpler alternatives.

The curriculum shows a clear effect on optimization. Without difficulty splitting, training directly on mixed sequence lengths of 1--1200 causes the loss to stall at 2.27, even with 980K training samples. In contrast, dividing the data into five progressively longer buckets yields stage-wise losses of 0.20, 0.28, 0.16, 0.02, and 0.034. Since the two settings use nearly identical training budgets---8.06B versus 8.1B tokens---the improvement comes from organizing samples by difficulty rather than simply increasing data volume. Weight inheritance is also critical. Without initializing each stage from the previous one, the training loss still decreases to 0.7, but the validation loss rises to 4.1. With inheritance, the training and validation losses decrease together to 0.6 and 0.8, respectively, indicating much more stable generalization.



\subsection{Online A/B Experiment}
\label{sec:ab}
\begin{figure*}[t]
  \centering
  \includegraphics[width=\textwidth]{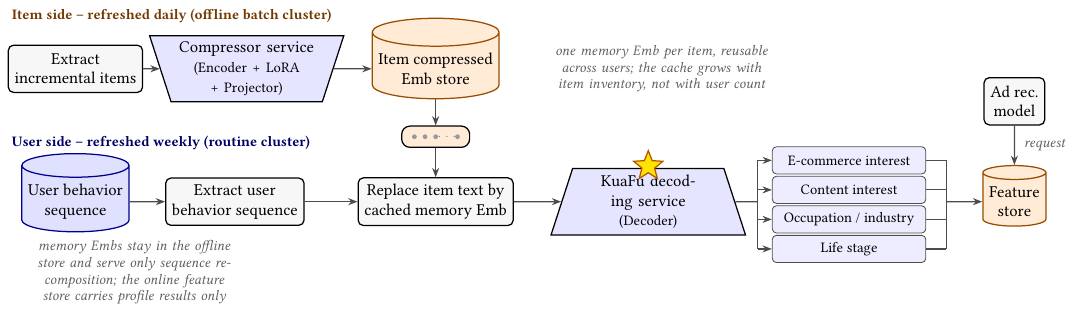}
  \caption{Production deployment of KuaFu: a daily item-side pipeline and a
  weekly user-side pipeline, served by two independently scaled clusters. The
  compressor is an offline batch, per-item cacheable computation; the decoder is
  a routine, per-user sequence-understanding computation. The offline cache grows with item inventory rather than with user count.
  }
  \label{fig:deploy}
\end{figure*}
We conduct an online A/B test with traffic randomly split by unique visitor. Both groups use the same resources for profile mining. The control group uses the previous profile features, which cover one year of behavior and are refreshed monthly, while the treatment group replaces them with KuaFu features covering two years of behavior and refreshed weekly. The recommendation model, training data, and all other features remain unchanged, isolating the effect of the user profile representation.

Traffic is gradually ramped from 1\% to 5\%, 20\%, 80\%, and finally 100\% over a four-week observation period. After full rollout, a 5\% holdout is retained for an additional ten months. Compared with this holdout, KuaFu achieves a 1.37\% increase in overall GMV, with a 95\% confidence interval of [0.71\%, 2.03\%].

Notably, both arms consume the same end-to-end resource budget, primarily GPU and storage. KuaFu uses the efficiency gains from compression to support a longer behavior horizon, increasing coverage from one year to two years, and a faster refresh cadence, from monthly to weekly. This allows longer-term preference signals to enter the recommendation model while enabling recent preference shifts to be reflected more quickly in online features.



\subsection{Production Deployment and Throughput}
\label{sec:deployment}

KuaFu is deployed on \platform{} to process two-year behavior histories for over one billion users. As shown in Figure~\ref{fig:deploy}, the production pipeline consists of two asynchronous flows. On the \emph{item side}, newly arrived items are processed daily by the compressor and stored in an item representation bank. Since each compressed representation depends only on the item itself, it can be cached once and reused across users and downstream tasks. On the \emph{user side}, profiles are refreshed weekly: item text in each user's behavior sequence is replaced with the corresponding cached representations, and the reconstructed sequence is fed to the KuaFu decoder to infer content interest, e-commerce interest, occupation/industry, and life stage. The resulting profiles are written to the online feature store for recommendation, while item representations remain in the offline representation bank.

The compressor and decoder are deployed on separate clusters because they have different workload characteristics. The compressor performs offline, per-item batch computation, whereas the decoder performs recurring, per-user sequence understanding. Separating the two allows them to scale independently and prevents decoder workloads from competing with item compression for resources. Together with item-level caching, this design makes the offline representation bank scale with the item inventory rather than with the number of users, enabling a weekly full refresh for over one billion users within the production latency and resource budget.

KuaFu substantially improves per-GPU throughput across downstream tasks. For life-stage profiling, QPM increases from 592 to 810 ((+37\%)), reducing GPU usage from 41 to 30. For e-commerce interest mining, QPM increases from 250 to 800 ((+220\%)), reducing GPU usage from 90 to 28. For content interest mining, QPM increases from 40 to 180 ((+350\%)), reducing GPU usage from 150 to 33. Overall, KuaFu saves \textbf{190 decoder-side GPUs}. Since the three tasks are provisioned independently according to their respective peak QPM requirements, the per-task savings are additive. Per-item storage costs and the total footprint for a deduplicated inventory of 20 billion items are reported in Appendix~\ref{app:ablation}.

\section{Conclusion and Future Work}
\label{sec:conclusion}

In this paper, we presented \textbf{KuaFu}, a unified framework for efficient user preference modeling over ultra-long behavioral sequences. KuaFu represents each behavioral item independently, applies two-axis compression to obtain compact and reusable memory representations, and assembles them chronologically for downstream user understanding. A progressive four-stage training pipeline improves information preservation and generation reliability, while hallucination-aware reinforcement learning further reduces unsupported outputs. KuaFu has been fully deployed on \platform{} for ten months, demonstrating substantial gains in both system efficiency and downstream business performance.


Several directions remain for future work. We currently use fixed compression ratios for different task families; adaptive, content-dependent compression may further improve the trade-off between efficiency and fidelity, particularly for sparse or highly heterogeneous behaviors. We also plan to explore richer compression schemes that jointly model token, feature, and temporal dimensions, as well as systematic scaling studies over model and data size. Finally, extending KuaFu to additional domains and continual-learning settings may enable more robust modeling of long-term preference evolution.

\section*{Statements}

\textbf{Reproducibility.}
Sections~\ref{sec:method} and~\ref{sec:exp} provide the model architecture, training objectives, optimization settings, data configuration, and ablation details. Proprietary behavior logs cannot be released for privacy reasons, while all public evaluation datasets and benchmarks are publicly available.

\textbf{Use of Generative AI.}
Generative AI was used only for language polishing and programming assistance in data processing and plotting. It was not used for research ideation, experimental design, or result generation, and all AI-assisted content was reviewed by the authors.

\textbf{Ethics.}
All user behavior logs are de-identified and processed in accordance with applicable privacy policies and regulations. No personally identifiable information is used or released, and cached representations are indexed by item rather than by user.

\textbf{Originality and Concurrent Submission.}
This manuscript is original and is not under review or accepted at another peer-reviewed venue with formal proceedings.

\bibliographystyle{ACM-Reference-Format}
\bibliography{refs_kuafu}

\appendix

\section{Training Recipe and Data Construction}
\label{app:recipe}

Table~\ref{tab:recipe} gives the full four-stage recipe behind
Table~\ref{tab:stages}. C/P/D denote the compressor, the projector and the
decoder; ``update'' and ``freeze'' refer to the trainable set of that stage.

\begin{table*}[t]
  \centering
  \caption{Four-stage training recipe and data statistics. C/P/D denote the
  compressor, the projector and the decoder; upd.\ / frz.\ mark whether the
  module is updated or frozen in that stage.}
  \label{tab:recipe}
  \footnotesize
  \setlength{\tabcolsep}{6pt}
  \begin{tabular}{llccclrr}
    \toprule
    Stage & Objective & C & P & D & Task composition & Samples & Tokens \\
    \midrule
    Pre-training         & restore text      & upd. & ---  & frz. &
      behavior seq.\ $+$ long text, bucketed by item count & 2.13M & 8.06B \\
    Post-tr.\ (simple)   & compressed QA     & upd. & upd. & frz. &
      8 behavior-understanding $+$ 7 generic NLP tasks     & 150K  & 0.80B \\
    Post-tr.\ (complex)  & compressed QA     & upd. & upd. & frz. &
      22 complex long-sequence $+$ 4 profiling tasks       & 350K  & 2.30B \\
    Joint-training          & fit both ends     & upd. & upd. & upd. &
      same composition as post-training                    & 350K  & 2.30B \\
    Hallu.-aware RL      & suppress 4 errors & frz. & frz. & upd. &
      long-sequence open-ended QA                          & 5K    & 41M   \\
    \bottomrule
  \end{tabular}
\end{table*}

\textbf{Understanding samples.} Understanding tasks are built
mainly from advertising behavior sequences. A large instruction model first
retrieves similar items to assist prompt writing; formal samples are then
generated in two steps: word frequencies are counted and nominal subjects
extracted to pin down the key information that must be preserved, and a
normalized question--answer pair is generated accordingly. Simple understanding
covers entity recognition, restating items at a given index, parity or
contiguous range, counting, and longest-interest-run extraction. Complex
understanding covers user profiling and interest mining, item similarity,
topic coverage and interest grouping, explanation of behavioral causes,
cross-source coordination, and temporal behavior analysis.

\begin{algorithm}[H]
\caption{KuaFu Batched Item Compression and Forward Pass}
\label{alg:kuafu}
{\footnotesize
\begin{flushleft}
\setlength{\parskip}{0.9pt}
\textbf{Input:} ordered items $[p_1,\dots,p_n]$ of user $u$, task prompt $q$,
target answer $y$, max item length $L_{\max}$, micro-batch size $B$, training
step $t$, total steps $T$\\
\textbf{Output:} cross-entropy loss $\mathcal{L}_{\mathrm{CE}}$\\[2pt]
\textbf{function} \textsc{Compress}($S,\mathit{Mask},t,T$):\\
\algI $\triangleright$ \textit{Compressor: encode each item into memory states}\\
\algI $\bar S,\overline{\mathit{Mask}} \gets \textsc{AppendMemory}(S,\mathit{Mask},[\mathrm{mem}_1,\dots,\mathrm{mem}_m])$\\
\algI $U \gets \textsc{TokenEmbed}(\bar S)$\\
\algI $U[\text{mem pos}] \gets \textsc{MemoryEmbed}([\mathrm{mem}_1,\dots,\mathrm{mem}_m])$\\
\algI $H \gets E(U,\overline{\mathit{Mask}})[\text{mem pos}]$\\
\algI $\triangleright$ \textit{Projector: compress token count and feature width}\\
\algI $\tilde Z \gets C\cdot\textsc{Down}(H)$\\
\algI $T_0 \gets T/3$\\
\algI $\alpha(t) \gets \tfrac{1}{2}\!\left[1+\cos(\pi t/T_0)\right]$ \textbf{if} $t\le T_0$ \textbf{else} $0$\\
\algI $Z \gets \textsc{Up}(\tilde Z)+\alpha(t)\cdot\textsc{Pool}(H)$\\
\algI \textbf{return} $Z$\\[2pt]
$S \gets \textsc{PadTruncateEach}([p_1,\dots,p_n],L_{\max})$\\
$\mathit{Mask} \gets \textsc{AttentionMask}(S)$\\
$Z_u \gets [\,]$\\
\textbf{for} $G$ \textbf{in} \textsc{MicroBatch}($(S,\mathit{Mask}),B$) \textbf{do}\\
\algI $Z_G \gets \textsc{Compress}(G.S,G.\mathit{Mask},t,T)$\\
\algI $Z_u \gets Z_u \mathbin{\Vert} Z_G$\\
\textbf{end for}\\
$S_{\mathrm{dec}} \gets \textsc{AssembleContext}(q,Z_u)$\\
$\hat y \gets D(S_{\mathrm{dec}})$\\
$\mathcal{L}_{\mathrm{CE}} \gets \textsc{CE}(\hat y,y)$\\
\textbf{return} $\mathcal{L}_{\mathrm{CE}}$
\end{flushleft}}
\end{algorithm}

\section{Full MRQA Results}
\label{app:mrqa}

Table~\ref{tab:mrqa-full} reports exact match and ROUGE-1 F1 at every
compression ratio, in-domain and out-of-domain, for every baseline.

\begin{table*}[t]
    \centering
    \caption{Full MRQA results, EM / ROUGE-1 F1 on a Llama-3.2-1B backbone. The
    out-of-domain columns use the same checkpoint, zero-shot.}
    \label{tab:mrqa-full}
    \footnotesize
    \setlength{\tabcolsep}{4pt}
    \begin{tabular}{lcccccc}
        \toprule
        & \multicolumn{3}{c}{In-domain} & \multicolumn{3}{c}{Out-of-domain} \\
        \cmidrule(lr){2-4}\cmidrule(lr){5-7}
        Method & $5\times$ & $15\times$ & $51\times$
        & $5\times$ & $15\times$ & $51\times$ \\
        \midrule
        ICAE~\cite{icae}
        & 33.82 / 47.53 & 31.28 / 44.5 & 27.99 / 40.39
        & 20.51 / 31.22 & 20.18 / 30.81 & 17.85 / 27.98 \\
        500xCompressor~\cite{x500compressor}
        & 40.14 / 55.26 & 34.98 / 49.35 & 30.26 / 43.19
        & 25.4 / 38.46 & 22.09 / 34.04 & 20.09 / 30.43 \\
        EPL~\cite{epl}
        & 46.33 / 62.9 & 36.65 / 51.52 & 30.26 / 43.22
        & 31.3 / 46.95 & 23.83 / 36.74 & 19.48 / 30.22 \\
        SAC~\cite{sac}
        & 46.95 / 63.63 & 39.67 / 54.95 & 33.08 / \textbf{46.37}
        & 32.3 / 47.72 & 26.02 / 39.26 & 21.44 / 32.24 \\
        \textbf{KuaFu (ours)}
        & \textbf{52.5 / 66.1} & \textbf{50.8 / 61.7} & \textbf{36.4} / 45.9
        & \textbf{50.0 / 60.1} & \textbf{43.1 / 54.2} & \textbf{37.5 / 48.2} \\
        \bottomrule
    \end{tabular}
\end{table*}

\section{Full RecBench Results}
\label{app:recbench}

\begin{figure*}[t]
  \centering
  \includegraphics{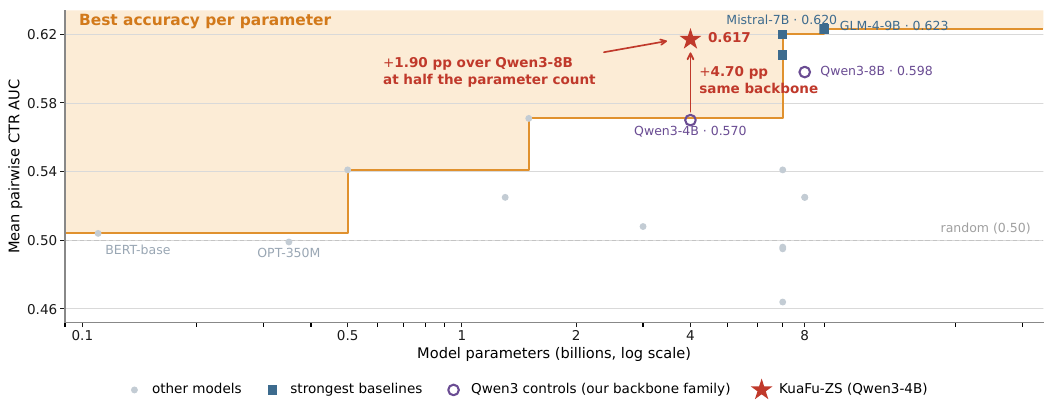}
  \caption{RecBench zero-shot pairwise CTR, mean AUC over the five datasets,
  against parameter count. Four notes. (i) GPT-3.5 (0.523) is served through an
  API and P5-Beauty (0.505) reports no parameter count, so neither is plotted;
  the two strongest models overall are Mistral-7B (0.6199) and GLM-4-9B
  (0.6231), both marginally above KuaFu-ZS. (ii) All differences in the body
  are computed on \emph{unrounded} means -- KuaFu-ZS is $0.6166$, drawn here as
  $0.617$ -- so recomputing them from three-decimal figures differs by
  $0.03$--$0.05$\,pp. (iii) The mean hides one regression: on the CDs subset
  the compressed representation brings no gain ($0.489$, below the $0.510$ of
  the Qwen3-4B backbone). (iv) KuaFu-ZS gets no task-specific training; only
  its user-behavior input is compressed.}
  \label{fig:recbench}
\end{figure*}

Figure~\ref{fig:recbench} places all twenty zero-shot models of
RecBench~\cite{recbench} on the accuracy-versus-size plane, together with the
caveats that the five-dataset mean alone does not convey.

\section{Ablation Details and Production Storage}
\label{app:ablation}

Table~\ref{tab:ab-full} collects the per-group numbers behind
Table~\ref{tab:ablation} together with the production storage footprint.

\begin{table*}[t]
  \centering
  \caption{Ablation details and production storage. (a) compressor curriculum:
  both settings consume comparable total tokens, so the difference is the
  difficulty organization alone. (b) inheriting pre-trained compressor weights
  before downstream SFT. (c) token-axis composition, width-axis strategy and
  decoder backbone. (d) storage for a deduplicated inventory of 20 billion
  items shared across users.}
  \label{tab:ab-full}
  \scriptsize
  \setlength{\tabcolsep}{4pt}
  \begin{minipage}[t]{0.48\textwidth}
    \centering
    \textbf{(a) Compressor curriculum}\\[2pt]
    \begin{tabular}{@{}llrrrr@{}}
      \toprule
      Setting & Items & Samples & Avg.\ tok. & Total tok. & Loss \\
      \midrule
      \multirow{5}{*}{curriculum}
        & 1        & 550K & 133  & 73.15M & 0.2 \\
        & 1--3     & 400K & 267  & 106.8M & 0.28 \\
        & 1--200   & 500K & 5330 & 2.6B   & 0.16 \\
        & 1--500   & 600K & 7656 & 4.6B   & \textbf{0.02} \\
        & 1--1200  & 83K  & 8202 & 680M   & \textbf{0.034} \\
      \midrule
      no curriculum & 1--1200 & 980K & 8245 & 8.1B & 2.27 \\
      \bottomrule
    \end{tabular}

    \medskip
    \textbf{(b) Inheritance}\\[2pt]
    \begin{tabular}{@{}lccc@{}}
      \toprule
      & init.\ loss & train loss & val.\ loss \\
      \midrule
      without & 3.2 & 0.7 & \textbf{4.1} \\
      with    & 1.2 & 0.6 & \textbf{0.8} \\
      \bottomrule
    \end{tabular}

    \medskip
    \textbf{(d) Production storage}\\[2pt]
    \begin{tabular}{@{}lcccc@{}}
      \toprule
      Compression & $20\times$ & $10\times$ & $2.5\times$ & $1\times$ \\
      \midrule
      (token, width) & (2, 128) & (2, 256) & (2, 1024) & (2, 2560) \\
      per item       & 512\,B   & 1\,KB    & 4\,KB     & 10\,KB \\
      total (TiB)    & 9.3      & 18.6     & 74.5      & 186.2 \\
      \bottomrule
    \end{tabular}
  \end{minipage}\hfill
  \begin{minipage}[t]{0.46\textwidth}
    \centering
    \textbf{(c) Projector axes and decoder backbone}\\[2pt]
    \begin{tabular}{@{}lr@{}}
      \toprule
      Token axis & train loss \\
      \midrule
      pooling                     & 1.9967 \\
      adapter                     & 1.8026 \\
      adapter $+$ pooled residual & \textbf{1.4148} \\
      \midrule
      Width axis & train reward \\
      \midrule
      naive                           & 0.1 \\
      $+$ progressive residual        & 0.25 \\
      $+$ differentiated LR annealing & 0.6 \\
      $+$ continued training          & \textbf{0.8} \\
      \midrule
      Decoder backbone & Base / SFT / DAPO \\
      \midrule
      life stage          & 0.74 / 0.81 / 0.81 \\
      e-commerce interest & 0.806 / 0.823 / \textbf{0.826} \\
      \bottomrule
    \end{tabular}
  \end{minipage}
\end{table*}

\section{Judging Protocol for Long-Sequence Open QA}
\label{app:judge}

Each answer is judged against the uncompressed original text along five
independent dimensions; the first four are scored present/absent and the fifth
on a three-level scale. In the RL reward the four error types are weighted by
the harm they do to the user: fabrication $0.45$, date misattribution $0.24$,
logic/task error $0.24$, missed detection $0.07$. The judge is instructed to be lenient about minor
wording and mild inference, and to answer ``present'' only for substantive
errors.

\begin{itemize}\itemsep1pt
  \item \textbf{Fabrication.} Introduces an entity, brand, person, place, title,
    organisation, number, event or direct quotation that is absent from the
    source; or invents a platform / region label the source never gives.
  \item \textbf{Missed detection.} Asserts ``none / not found / no such $X$''
    while the source contains at least one relevant entry, including synonymous
    or near-synonymous phrasings.
  \item \textbf{Date misattribution.} Assigns an event to the wrong date,
    misquotes a year / month / day, or invents a period label the source never
    gives.
  \item \textbf{Broken logic.} Factual distortion other than dates -- a value
    reversed in direction, an attribute assigned to the wrong subject, a broken
    correspondence, self-contradiction, or a claim that conflicts with an
    explicit fact; or reversed causality and plainly invalid inference.
  \item \textbf{Completeness.} Complete / partial / missing, with respect to the
    parts of the instruction that the answer actually addresses.
\end{itemize}

\paragraph{Robustness to the choice of judge.}
The hallucination results in Section~\ref{sec:exp} are reported under a single
automatic judge, which raises the question of whether they are an artifact of
that judge. To rule this out we re-measured the six headline metrics under three
independent protocols: human annotation on a held-out subset of the evaluation
set, the primary test judge GPT-5.5, and a second strong judge Opus-4.8.
Table~\ref{tab:judge-agree} reports the DAPO\,$-$\,joint-training change of each
metric under all three, together with the mean and standard deviation across the
two LLM judges and the Cohen's $\kappa$ agreement of each LLM judge with the
human labels. All three protocols agree on the sign and, within $1$\,pp, the
magnitude of every metric; $\kappa$ ranges from $0.72$ to $0.91$
(substantial-to-almost-perfect agreement). In particular, the asymmetry the
reward is designed to induce -- fabrication crushed while missed detection is
allowed to drift up -- is reproduced by the human raters, so it reflects the
model rather than the judge.

\begin{table*}[t]
  \centering
  \caption{Judge robustness for the hallucination-aware RL results. Each cell is
  the DAPO\,$-$\,joint-training change of the metric in percentage points, measured
  under human annotation, the primary test judge GPT-5.5, and a second judge
  Opus-4.8. ``LLM mean $\pm$ SD'' is taken over the two LLM judges; ``$\kappa$''
  is Cohen's agreement of each LLM judge with the human labels. The GPT-5.5
  column reproduces exactly the values drawn in Figure~\ref{fig:rl}.}
  \label{tab:judge-agree}
  \footnotesize
  \setlength{\tabcolsep}{6pt}
  \begin{tabular}{@{}lrrrcc@{}}
    \toprule
    Metric (DAPO $-$ joint-training) & Human & GPT-5.5 & Opus-4.8 & LLM mean $\pm$ SD & $\kappa$ (GPT\,/\,Opus) \\
    \midrule
    Fabrication hit rate $\downarrow$         & $-28.4$ & $-29.0$ & $-27.8$ & $-28.4 \pm 0.8$ & 0.89 / 0.84 \\
    Missed detection hit rate $\downarrow$    & $+1.3$  & $+1.5$  & $+1.7$  & $+1.6 \pm 0.1$  & 0.79 / 0.72 \\
    Date misattribution hit rate $\downarrow$ & $-8.1$  & $-8.5$  & $-8.8$  & $-8.7 \pm 0.2$  & 0.86 / 0.80 \\
    Logic / task error hit rate $\downarrow$  & $-5.2$  & $-5.5$  & $-5.1$  & $-5.3 \pm 0.3$  & 0.83 / 0.76 \\
    Fully correct answers $\uparrow$          & $+33.0$ & $+34.0$ & $+32.6$ & $+33.3 \pm 1.0$ & 0.91 / 0.86 \\
    Complete answers $\uparrow$               & $+0.7$  & $+0.5$  & $+0.9$  & $+0.7 \pm 0.3$  & 0.81 / 0.74 \\
    \bottomrule
  \end{tabular}
\end{table*}

\section{Difficulty-Increasing Curriculum: Task Families}
\label{app:curriculum-tasks}

\emph{Needle in a haystack.} For a given item in the sequence, find every
related item; given a target type, find every item of that type; run entity
recognition over all items; summarize all items; find all items at odd
positions; find all items at even positions; return the items in positions
$j$ through $k$; return the items at several randomly specified positions.

\emph{Interest distribution and diversity.} Given the behavior sequence
(a smaller index means a more recent behavior), mine sub-category interests
under each main interest category. The output is JSON with a
\texttt{subcategory\_interests} dictionary, keyed by main category, whose
values are lists of sub-categories carrying a keyword, a count and supporting
evidence, plus a short \texttt{reasoning} field.

Note on data handling: the evidence field holds de-identified references
rather than verbatim behavior text, consistent with the ethics statement in
the body.

\end{document}